\documentclass{PoS}

\usepackage{wrapfig}

\usepackage[left]{lineno}

\title{Status and Prospects for Heavy Flavour Physics at LHC}

\ShortTitle{Status and Prospects for Heavy Flavour Physics at LHC}

\author{\speaker{Pascal Perret}\thanks{On behalf of the ATLAS, CMS and LHCb Collaborations.}\\
        Clermont Universit\'e - CNRS/IN2P3 - Laboratoire de Physique Corpusculaire - BP 80026 - \\ F 63177 Aubi\`ere Cedex - France\\
        E-mail: \email{perret@in2p3.fr}}

\abstract{
The Large Hadron Collider will be a unique place to find new physics in the next decade. A huge production of b and c quarks will allow a rich programme of Heavy Flavour Physics to be carried out either by the multipurpose experiments ATLAS and CMS or by LHCb, the experiment designed  for such physics. An overview of the LHC machine and experiments' performances will be given with the first 2010 data. The start-up is very bright and some first LHC heavy flavour results will be presented. The b physics program at LHC will be illustrated with three examples: the searches for rare decays such as $B_s \to \mu\mu$, the CP measurements from $B_s \to J/\psi \phi$ and CP measurements of the Cabibbo Kobayashi Maskawa (CKM) angle $\gamma$. Some prospects for an upgrade  of the LHCb detector will also be given.   
}

\FullConference{The Xth Nicola Cabibbo International Conference on Heavy Quarks and Leptons,\\
		October 11-15, 2010\\
		Frascati (Rome) Italy}

\begin{document}


\section{Introduction}
\vspace{-1.07mm}
Beside the direct search for new particles, high precision measurements in the b and c sector are a complementary way to find New Physics (NP). After many successes in heavy flavour physics over the past decade, conducted by the B factory experiments Babar and Belle, as well as the CDF and D0 experiments at the Tevaton hadronic collider, the proton-proton Large Hadronic Collider (LHC) installed at CERN will be the most intense source of b and c hadrons. Even though it has been confirmed that the CKM mechanism is the major source of CP violation observed so far and that the description in the Standard Model (SM) of flavour changing neutral current is correct, there is still room for NP. The accuracy of the tests is still limited. The $b \to d$  transition has been measured at the level of 10-20\% accuracy, however NP effects can still be large in other processes like the $b \to s$ transition. 

LHC experiments will have a rich heavy flavour physics program thanks to a design LHC luminosity of $10^{34}$ cm$^{-2}$s$^{-1}$ and a large $b \bar b$ cross section of $\sim$ 500 $\mu$b at a centre-of-mass energy of  $\sqrt{s}$= 14 TeV, which represents about 1\% of all visible interactions. As in the case of the Tevatron, all flavours of $b$-hadrons will be produced ($B_{d}$, $B_{u}$, $B_{s}$, $B_{c}$, $\Lambda_{b}$, ...). At design parameters, the LHC will collide proton bunches at a frequency of 40 MHz. Three of the LHC experiments will study CP violation and rare decays in the $b$ sector: the two general purpose detectors ATLAS and CMS, and the LHCb detector dedicated to heavy flavour physics. 
ATLAS and CMS are hermetic and have a full coverage up to a pseudo rapidity of 2.5. They have been designed to run at the full LHC luminosity of  $10^{34}$ cm$^{-2}$s$^{-1}$ for the direct search of new particles. Nevertheless they have a $b$ physics program in the first years of LHC when running at a luminosity of $10^{33}$ cm$^{-2}$s$^{-1}$, when it is planed to collect around 10 fb$^{-1}$ per year. Both experiments have  first a fast custom-built electronics trigger which reduces the event rate from 40 MHz to below 100kHz, exploiting mainly muon detectors. Software triggers reduce the rate to an output rate of around 200Hz, $b$ physics being accounted for 5 to 10\% of total trigger resources.   Beyond this luminosity first the number of interactions per crossing can reach 10 or 20 and is not appropriate for precision measurements, second triggering is even more challenging and only some specific rare $b$ decays with two muons in the final  states can still be studied. 

On the contrary LHCb is an optimized detector for $b$ and $c$ physics precision studies at LHC. It is a single-arm open spectrometer, covering the pseudo rapidity range: $[1.9 - 4.9]$, in order to exploit the fact that the $b \bar b$ pair production is sharply peaked forward-backward. It incorporates precision vertexing and tracking systems, particle identification over a wide momentum spectrum and the capability to trigger down to very small momenta. LHCb will operate at a reduced luminosity of 2 $\times$ $10^{32}$cm$^{-2}$s$^{-1}$ which will be kept locally controlled by appropriately focusing the beam. At this luminosity the majority of the events have a single $pp$ interaction. The fast custom-built electronics trigger of LHCb will reduce the trigger rate only to 1 MHz by applying much lower p$_{\rm T}$ / E$_{\rm T}$ trigger thresholds on muon, electron, photon and hadron candidates. The remaining reduction is performed by software triggers that exploit the full detector readout and lead to an output rate of 2 kHz of interesting heavy flavour events across a wide spectrum of final states.

\section{LHC machine and experiments' performance}
\resetlinenumber[1]
A lot of records have been reached by the LHC in 2010. First, on 30 March 2010 beams collided for the first time at $\sqrt{s}$=7 TeV in the LHC, marking the start of the LHC research programme. 
The first collisions were done with one bunch per beam, corresponding to a luminosity of $10^{27}$ cm$^{-2}$s$^{-1}$, far from the ultimate goal of 2808 bunches. 
The number of particles in each bunch was increased, the beam size at the interaction point was squeezed down, then the machine ran with 13 bunches in each beam, allowing in May
to set a new luminosity record of 2 $\times$ $10^{29}$ cm$^{-2}$s$^{-1}$. 
%
LHC was able to run smoothly with bunches at the design intensity, that is, with 1.1 $\times$ $10^{11}$ protons per bunch. A big jump in luminosity was obtained mid September when fills where delivered with 56 bunches arranged in trains of eight bunches per train,
47 bunches colliding at ATLAS, CMS and LHCb. 
The number of bunches and trains was increased and, on October 14$^{\rm th}$, the main objective of reaching a luminosity of $10^{32}$cm$^{-2}$s$^{-1}$ by the end of 2010 proton running was reached. 
On November 4$^{\rm th}$, the proton running for 2010 in the LHC came to the end. 
Physics running was interspersed with periods of machine development and around 40 pb$^{-1}$ have been collected by ATLAS, CMS and LHCb. The experiments have been running with about 90\% efficiency. During the year the trigger of the detectors, which are highly configurable, evolved a lot to match the exponential increase of the LHC luminosity 
and the physics requirements.
The next target for the machine is to collect an integrated luminosity of 1 fb$^{-1}$ before the end of 2011.

The detectors are performing very well. The calorimeters were among the first detectors to be calibrated. Quickly a mass resolution on the $\pi ^0$ mass of 20 MeV/c$^2$ for ATLAS, 14 MeV/c$^2$ for CMS and 7.2 MeV/c$^2$ for LHCb was obtained, close to Monte Carlo (MC) expectations. The impact parameter resolution for 2 GeV/c tracks is 60 $\mu$m in ATLAS, 50 $\mu$m in CMS and 25 $\mu$m in LHCb, not far from MC expectations. Further improvement will be achieved by better alignment. In the LHCb VErtex LOcator detector the alignment among sensors is better than 4 $\mu$m and the fill-to-fill variations are less than this value. Over a wide range the di-muon spectrum has been measured. The $J/\psi \to \mu \mu$ events have been used as a standard candle for calibration of the tracking system. In the whole detector acceptance a mass resolution on $J/\psi$ (resp. $\Upsilon$)  of 70 (170) MeV/c$^2$ for ATLAS, 47 (100) MeV/c$^2$ for CMS and 14 (47) MeV/c$^2$ for LHCb was obtained, approaching MC expectations. Particle identification is already performing amazingly well. For instance, LHCb which has a very powerful particle identification system over a wide momentum range (2-100 GeV/c) based on two Ring Imaging CHerenkov detectors, obtained an average efficiency for $K^\pm$ detection of about 95\% with a $\pi \to K$ misidentification probability of 7\%, close to MC level performance.

\section{First LHC heavy flavour results}
\resetlinenumber[1]
Clean charm signals reconstructed in the first nb$^{-1}$ of data~\cite{spectro, LHCbspectro} already allow to firm up exciting prospects for measurements of $D^0 - \bar{D}^0$ mixing and CP violation in the charm sector~\cite{LHCbsearch}. Bottom production has been measured with a few nb$^{-1}$ of data at $\sqrt{s}$= 7 TeV using the inclusive decay $b \to J/\psi X$~\cite{LHCbspectro, bprod}. In addition LHCb used a second method~\cite{LHCbprod} which exploits the semileptonic channel $b \to D^0 \mu \nu X$, with $D^0 \to K^- \pi^+$. In this analysis the $D^0$ mesons from a parent $b$ are separated from those directly produced on the basis of the impact parameter of the reconstructed $D$ tracks measured with respect to the primary vertex. Integrating over the LHCb pseudo rapidity acceptance yields the value 
$	\sigma_b(pp \to H_b X; \sqrt{s}=7 \, {\rm TeV}; 2 < \eta_b < 6) = 75.3 \pm 5.4 \pm 13 \; \mu \rm b $.
 Extrapolating to the full solid angle using PYTHIA 6.4 and averaging with the LHCb preliminary measurement obtained with the first method corresponds to a total $b\bar b$ production cross-section $\sigma (b \bar b; \sqrt{s} $= 7 TeV) = 298 $\pm$ 15 $\pm$ 43 $\mu \rm b$~\cite{LHCbspectro, LHCbsemi}. This value agrees with the expectations and confirms that the $b$ yield assumed in the design of LHCb was correct.


\section{Future for Heavy Flavour Physics at LHC}

\subsection{New physics in $B^0_s \to \mu^+\mu^-$}
\resetlinenumber[1]
The $B^0_s \to \mu^+\mu^-$ decay is predicted to be very rare in the SM (BR $ = (3.6 \pm 0.3) \times 10^{-9}$~\cite{buras}) since it involves flavour-changing neutral currents and experiences a large helicity suppression, but is sensitive to NP and could be strongly enhanced in SUSY. The best current limit is achieved by CDF (BR $ < 3.6 \times 10^{-8} \, @ \, 90\%$ CL)~\cite{CDF_bsmu} while D0 achieved BR $ < 4.2 \times 10^{-8} \,  @ \, 90\%$ CL~\cite{D0_bsmu}. The final state of this mode containing only muons it is easily accessible to ATLAS, CMS and LHCb. This experimental search has to deal with the problem of an enormous level of background, dominated by random combinations of two muons originating from two distinct $B$ decays. Since an absolute branching ratio measurement would be experimentally challenging, it is planed to measure the branching ratio of this decay relative to the control channel $B^+ \to J/\psi K^+$ which was already observed in the first LHC data. 
ATLAS and CMS plan to perform cut based analyses to separate signal from background using similar variables~\cite{ATLASrare, CMSrare}. Assuming $\sigma_{b \bar b}$ = 500 $\mu b$ $@ \sqrt{s}$= 14 TeV, with 10 fb$^{-1}$ ATLAS expect to have 5.6 signal candidates for 14 background candidates, while CMS with 1 fb$^{-1}$ expect to have 2.4 signal candidates for 6.5 background candidates. This last number can be translated in an exclusion limit of  BR $ < 1.6 \times 10^{-8} \, @ \, 90\%$ CL which corresponds to a BR $< 2.1 \times 10^{-8} \, @  \, 90\%$ CL by rescaling with the quoted measurement of $\sigma_{b \bar b}$ by LHCb at $\sqrt{s}$= 7 TeV. Limited amount of MC is a cause of large uncertainty on these estimations.


LHCb will exploit its excellent tracking and vertexing capabilities and will use an approach to measure this branching ratio which is philosophically similar to Tevatron. A loose selection will be applied and then a global likelihood will be constructed, the analysis being made in a 3-parameter space~\cite{LHCbrare}. Prospects from the data are  encouraging. The two body hadronic decays, $B \to h^+ h^-$ are used for control. A mass resolution of 24 MeV/c$^2$ has been measured, very close to MC expectations (22 MeV/c$^2$). The Impact Parameter resolution is in agreement with MC for p$_{\rm T} >$ 2 GeV and the background is at the expected level. With 1 fb$^{-1}$ LHCb expects 6 signal candidates for 30 background candidates, with the measured $\sigma_{b \bar b}$ at $\sqrt{s}$= 7 TeV. 
This yields a sensitivity to exclude a branching ratio down to 3.4 $\times 10^{-8} \,  @  \, 90\%$ CL with 50 pb$^{-1}$ (close to the data set collected by the end of 2010), approaching Tevatron's limit.
The data set collected by the end of 2011 should be 1 fb$^{-1}$. It
 should allow the observation by LHCb of this decay at 5$\sigma$ down to 5 times the expectations from the SM  (i.e. for BR $ > 1.7 \times 10^{-8}$) or to exclude branching ratios down to 7 $\times 10^{-9}   @  90\%$ CL.

\subsection{$\beta_s$ measurements from $B_s \to J/\psi \phi$}
\label{sect:phys}
\resetlinenumber[1]

The interference between $B^0_{s}$ decay to $J/\psi \phi$ with or without mixing gives rise to a CP violating phase $\phi_{J/\psi \phi}$. In the SM this phase is very small and predicted to be -2$\beta_s \simeq$-0.04, where $\beta_s$ is the smallest angle of the "b-s unitarity triangle". But it can receive sizable
NP contributions through box diagrams. CDF and D0 have reported measurements of the $B_s$ mixing phase with a large central value of $\beta_s$, but with a poor precision~\cite{CDFD0betas}.  

\begin{wrapfigure}{r}{60mm}
  \centering
  \includegraphics[width=2.5 in]{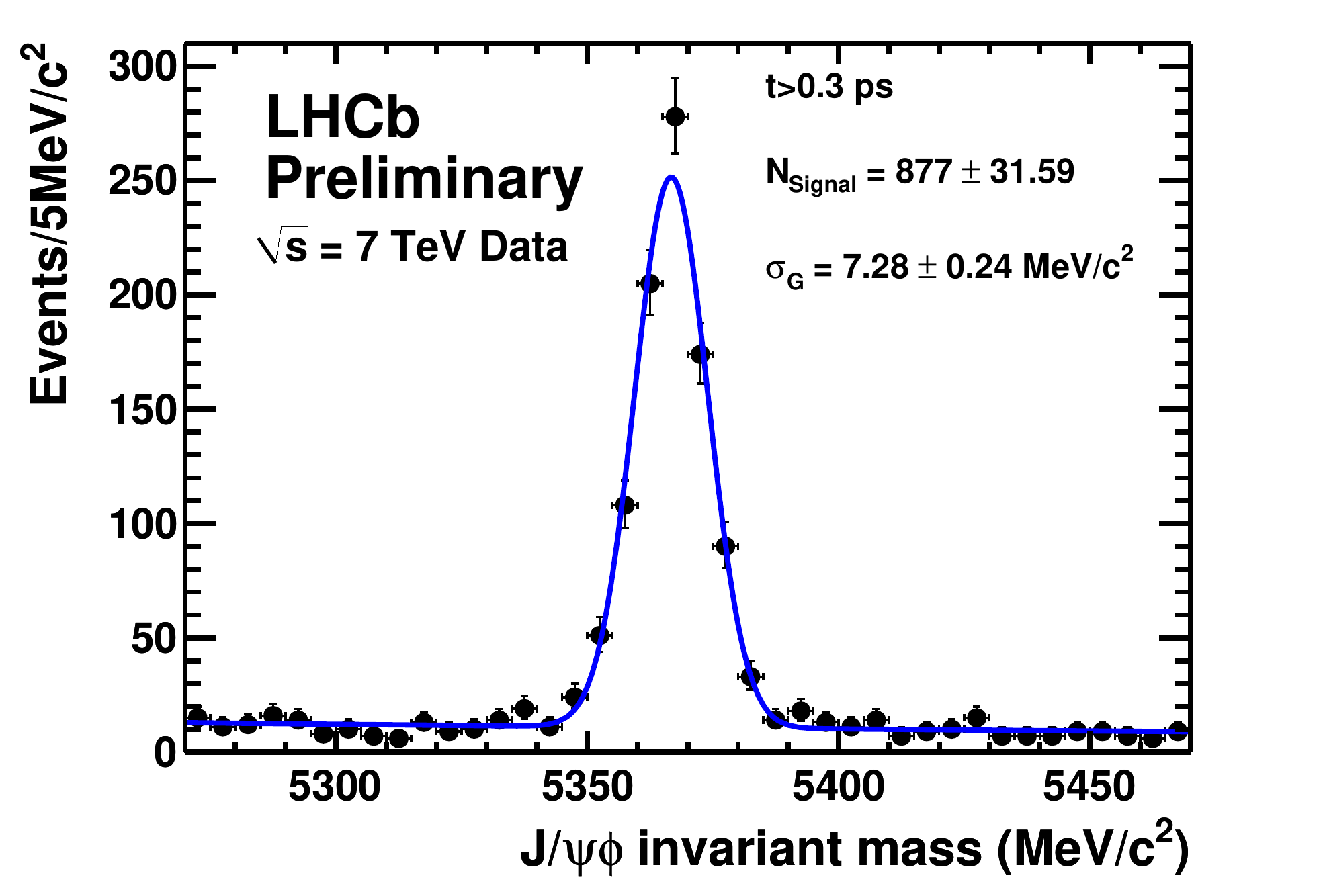}
  \caption{\small LHCb $B^0_{s} \to J/\psi \phi$ candidates in the first 34 pb$^{-1}$  after applying a $J/\psi$ mass constraint and a cut on the lifetime. }
  \label{Fig_LHCbbsphi}
\end{wrapfigure}

This analysis is complicated by the fact that the final state involves two vector mesons and so two orbital
momentum states can occur. Determination of the CP violation parameter as well as the contributions from the different angular momentum states is achieved by a time-dependent, flavour-tagged, angular analysis of the decay rates. All LHC experiments have access to this decay as they can trigger on the $J/\psi$ decaying into a muon pair, but ATLAS and CMS studies on this topics~\cite{ATCMphis} are quite old and need to be revisited, so prospectives in the following will refer to LHCb only. 


With 2 fb$^{-1}$ of data taking, LHCb should reconstruct 114k events, with S/B=2. Note that the LHCb background levels are expected to be significantly higher because LHCb uses a lifetime unbiased event selection which results in a signal sample with a higher per event sensitivity to the parameter $\phi_{J/\psi \phi}$.  This will allow a precision of 0.03 rad on $\phi_{J/\psi \phi}$. First data are encouraging~\cite{LHCbsearch}. With the first 34 pb$^{-1}$ of data taking, LHCb has reconstructed around one thousand of $B_s \to J/\psi \phi$ candidates (Figure~\ref{Fig_LHCbbsphi}). Already at the start-up with $\cal O$(0.1 fb$^{-1}$), LHCb should rapidly achieve better sensitivity than Tevatron and pin down whether there really is any sign of new physics.

\subsection{$\gamma$ measurements}
\resetlinenumber[1]
The CKM angle $\gamma$ is the least well constrained of the angles of the unitarity triangle. Many channels can contribute to $\gamma$ measurements. The $B^0_{d, s} \to h^+ h^-$ family of decays, where $h$ stands for a $\pi$ or $K$ meson, have decay rates with non negligible contributions from penguin diagrams, making them sensitive to NP. The dependence on $\gamma$ comes from CP time-dependent measurements from $B^0_{d} \to \pi^+ \pi^-$ and $B^0_{s} \to K^+ K^-$ which allow to extract $\gamma$ relying on U-Spin symmetry. The decay $B^0_s \to D^\pm_s K^\mp$ and its charge conjugate can
proceed through two tree  decay diagrams, the interference of which
gives access to the phase $\gamma$ 
if the $B_s \bar{B_s}$ mixing phase is determined otherwise ($e.g.$ with 
$B_s \to J/\psi \phi$, see section~\ref{sect:phys}). The charm decays of charged B mesons proceed through tree level diagrams and enable a direct SM measurement of $\gamma$. Different strategies exist for measuring $\gamma$, depending on the final states. Because all the involved decays used to measure $\gamma$ have fully hadronic final states, ATLAS and CMS will not be able to contribute to these measurements, but LHCb will greatly improve the precision on $\gamma$. Crucial use of particle identification thanks to the RICH detectors and very good mass resolution will be the key ingredients for LHCb measurements. 
In the first sample of $\sim$ 3.1 pb$^{-1}$ data taking LHCb has observed nice $B^0_{d, s} \to h^+ h^-$ peaks, with a yield matching so far the expectations. In 2011 running LHCb will get largest world samples both in $B^0$ and $B_s$.  It is estimated~\cite{LHCbgamma} that the LHCb achievable precision on $\gamma$ will be (4-5)$^\circ$ with 2 fb$^{-1}$ of data taking.

\section{Prospects for Heavy Flavour Physics at LHC: LHCb upgrade}
\resetlinenumber[1]
Within few years we will need even more precisions and so statistics either to understand the nature of NP that we will have found at LHC and to elucidate its flavour structure or to probe higher mass scale if NP is not yet seen ... In both case it is very important to be able to run the LHCb detector at higher luminosity, LHC being a natural  b factory, in order to collect around 100 fb$^{-1}$. The LHCb upgrade strategy consists to run at 10 times its design luminosity (i.e. 2 $\times$ $10^{33}$ cm$^{-2}$s$^{-1}$). 
A first phase, around 2016 will allow to run at 
$10^{33}$ cm$^{-2}$s$^{-1}$.
In order to increase efficiency, the trigger strategy has to be redesigned and based on software only allowing to use many discriminants including secondary vertex information. This is possible with a readout of the whole detector at 40 MHz. So the first trigger level is removed and High Level Trigger will reduce the output rate down to 20 kHz, a factor 10 higher than at present. This software trigger will be very flexible and able to cope with any scenarii.    
As a consequence, the front-end electronics have to be rebuilt in order to deal with the 40 MHz readout. This is
also the case of some very front-end ASIC which are limited to a readout frequency of 1 MHz. In addition, an increasing of the luminosity will cause higher occupancy and higher radiation dose for the detectors
around the beam pipe. Therefore, the inner parts of the tracking system have to be redesigned, namely the vertex
detector, the trigger tracker and the inner tracker. This is the baseline of the first phase of the LHCb upgrade and R\&D has started in many areas in order to achieve it. 
A Letter Of Intent will be published early in 2011. This upgrade will be complementary to the one for a Super B factory (SuperKEKB/Belle II or SuperB).


\end{document}